\documentclass[a4paper,onecolumn,11pt]{IEEEtran}

\usepackage{amsmath,amssymb}
\usepackage{amsthm}

\usepackage{graphicx}
\usepackage{tikzexternal}
\allowdisplaybreaks

\newcommand{\F}{\mathbb{F}}

\newcommand{\ga}{\alpha}
\newcommand{\gb}{\beta}

\newcommand{\gd}{\delta}

\newcommand{\gw}{\omega}
\renewcommand{\phi}{\varphi}

\newcommand{\calP}{\mathcal{P}}

\newcommand{\frakm}{\mathfrak{m}}

\newcommand{\nn}[1]{\makebox[1.5em]{$#1$}}

\newtheorem{thm}{Theorem}
\newtheorem{lem}[thm]{Lemma}
\newtheorem{prop}[thm]{Proposition}

\newcommand{\set}[1]{\{#1\}}

\newcommand{\ideal}[1]{\langle{#1}\rangle}
\newcommand{\ev}{\mathrm{ev}}
\newcommand{\wt}{\mathrm{wt}}

\DeclareMathOperator{\LT}{lt}
\DeclareMathOperator{\LC}{lc}

\begin{document}

\title{Unique Decoding of Plane AG Codes \\ via Interpolation}

\author{Kwankyu Lee, Maria Bras-Amor\'os, and Michael E.~O'Sullivan
\thanks{K.~Lee is with the Department of Mathematics, Chosun University, Gwangju 501-759, Korea (e-mail: kwankyu@chosun.ac.kr). His work is supported by Basic Science Research Program through the National Research Foundation of Korea(NRF) funded by the Ministry of Education, Science and Technology(2009-0064770) and also by research fund from Chosun University, 2008.}%
\thanks{M.~Bras-Amor\'os is with the Department of Computer Engineering and Mathematics, Universitat Rovira i Virgili, Tarragona 43007, Catalonia, Spain (e-mail: maria.bras@urv.cat). Her work is supported by the Spanish Government through the projects TIN2009-11689 ``RIPUP" and CSD2007-00004 ``ARES".}%
\thanks{M.~E.~O'Sullivan is with the Department of Mathematics and Statistics, San Diego State University, San Diego, CA 92182-7720, USA (e-mail: mosulliv@math.sdsu.edu). His work is supported by the National Science Foundation under Grant No. CCF-0916492.}%
}

\maketitle

\begin{abstract}
We present a unique decoding algorithm of algebraic geometry codes on plane curves, Hermitian codes in particular, from an interpolation point of view. The algorithm successfully corrects errors of weight up to half of the order bound on the minimum distance of the AG code. The decoding algorithm is the first to combine some features of the interpolation based list decoding with the performance of the syndrome decoding with majority voting scheme. The regular structure of the algorithm allows a straightforward parallel implementation.
\end{abstract}

\IEEEpeerreviewmaketitle


\section{Introduction}

Unique decoding of algebraic geometry codes is now a classical subject. By the works of Justesen et al., Skorobogatov and Vl{\u{a}}du{\c{t}}, and many others, the paradigm of decoding via syndromes using error locator polynomials and evaluator polynomials is well established \cite{hoholdt1995}. Enhanced by Feng and Rao's majority voting, the syndrome decoding algorithm for AG codes is capable of correcting errors up to half of the Feng-Rao bound, also called the order bound, which is no less than the designed distance. Until the advent of Guruswami and Sudan's list decoding algorithm based on interpolation \cite{guruswami1999}, the syndrome decoding algorithm had long been a uniquely available algorithm for decoding AG codes. Since then, the superiority in decoding performance of the list decoding algorithm has somewhat faded the syndrome decoding algorithm. 

The superior performance of the list decoding is gained at the expense of large computational complexity. Table \ref{nskqee} below, excerpted from \cite{kwankyu2009}, shows an experimental result about the performance of the list decoding algorithm for the Hermitian code of length $27$ and dimension $14$. Here $\tau$ denotes the number of errors that the list decoder is guaranteed to correct with multiplicity parameter $m$, and the number of successful decodings was counted out of 10,000 random error vectors of weight $t$. The notation $\infty$ is used when successful decoding is guaranteed, as $t\le\tau$. We may compare the result with the decoding performance of the syndrome decoding algorithm, which can correct errors of weight half of the designed distance, that is, 5 in this case. The list decoding algorithm certainly has better performance because, with multiplicity parameter 25,  it can decode up to 6 errors, though the increased complexity is prohibitively high. 

\begin{table}[ht]
\centering
\begin{minipage}{0.59\textwidth}
\[
\begin{array}{|c|c*{10}{|c}}
\hline
 & & \multicolumn{8}{c|}{t}\\
\cline{3-10}
m & \tau & \nn{2} & \nn{3} & \nn{4} & \nn{5} & \nn{6} & \nn{7} & \nn{8} & \nn{9} \\\hline
1 & 2    & \infty & 10000 & 10000 & 9977 & 998 & 85 & 2 & 0 \\
2 & 3    & \infty & \infty & 10000 & 10000 & 282 & 1 & 0 & 0 \\
3 & 4    & \infty & \infty & \infty & 10000 & 109 & 0 & 0 & 0 \\
5 & 5    & \infty & \infty & \infty & \infty & 1119 & 0 & 0 & 0 \\
25 & 6    & \infty & \infty & \infty & \infty & \infty &  &  &  \\\hline
\end{array}
\]
\end{minipage}
\caption{}\label{nskqee}
\end{table}

Moreover note that, to match the performance of the syndrome decoding algorithm, that is, to be guaranteed for successful decoding up to 5 errors, the multiplicity parameter should be at least 5. This means, for the same performance, the list decoding algorithm suffers slow decoding speed.  On the other hand, observe from the experiment that list decoding with multiplicity $m=1$ performs almost as well as syndrome decoding. It corrects most cases of 5 errors, but unfortunately misses some. Since successful decoding only up to 2 errors is guaranteed by the theory of list decoding, this is a much better performance than expected. 

Beside the performance, the two kinds of decoding algorithms, one based on interpolation and the other on syndromes, have different features. The list decoding algorithm decodes in the primal AG code, whose codeword is obtained by evaluation at rational points of the base curve, while the syndrome decoding algorithm decodes in the dual code. The former computes the message directly from the so-called $Q$-polynomial, while the latter obtains the message after computing the error locations and the error values from the error locator and evaluator polynomials. Finally, the syndrome decoding algorithm is equipped with the majority-voting scheme while there is no corresponding mechanism for list decoding.

These observations lead to the view, already widely accepted to experts in this area, that the list decoder with multiplicity one is closely related to the syndrome decoding algorithm without majority voting enhancement. However, they cannot be equivalent, principally due to the fact that one algorithm is for the primal code while the other one is for the dual code.  Hence there is a missing idea corresponding to the majority voting in the context of interpolation based decoding, to match up the performance of the syndrome decoding algorithm.  In this paper, we present an interpolation based unique decoding algorithm capable of correcting up to half of the order bound. The algorithm is an amalgamation of the decoding algorithm with multiplicity one and list size one in \cite{kwankyu2009} and a recursion procedure that resembles the majority voting of Duursma \cite{duursma1993}. Like list decoding, our unique decoding algorithm decodes in the primal codes and computes the message directly from the received vector. Like K\"otter's algorithm \cite{koetter1998}, it allows an efficient parallel implementation.

In Section \ref{sec_dwkqd}, we review basic concepts and establish notations regarding AG codes on plane algebraic curves. We refer the reader to \cite{fulton1969,stichtenoth2009,kunz2005} for the basic theory of algebraic curves and AG codes over finite fields, and \cite{cox1997a,atiyah1969} for Gr\"obner bases and commutative algebra. In Section \ref{sec_kdqdd}, we present and prove a unique decoding algorithm, using a majority voting procedure as a fundamental decoding method. In Section \ref{sec_dkdwas}, we give a decoding example of Hermitian codes. In Section \ref{sec_dwkqqw}, we conclude with brief remarks.

\section{Preliminaries}\label{sec_dwkqd}

Let $X$ be an irreducible plane curve defined by the equation $E(x,y)=0$ over a field $\F$ where 
\[
	E(x,y)=y^a+\sum_{ai+bj<ab}c_{i,j}x^iy^j+cx^b
\]
with $\gcd(a,b)=1$ and $0\neq c\in\F$. These curves are known as Miura-Kamiya curves in the literature \cite{sakata1995}. It is well known that $X$ has a unique point $P_\infty$ at infinity that is either nonsingular or a cusp. Hence there is a unique valuation $v_{P_\infty}$ associated with $P_\infty$. Let $\gd(f)=-v_{P_\infty}(f)$ for $f$ in the coordinate ring $R$ of $X$. Let $\gd_x=\gd(x)=a$ and $\gd_y=\gd(y)=b$. By the equation of the curve, the ring $R=\F[x,y]$ is a free module over $\F[x]$ of rank $a$ with basis $\set{y^j\mid 0\le j<a}$. The semigroup of R at $P_\infty$
\[
	S=\set{\gd(f)\mid f\in R}=\set{i\gd_x+j\gd_y\mid 0\le i,0\le j<a}
\]
is a subset of the Weierstrass semigroup at $P_\infty$. For each nongap $s\in S$, there is a unique monomial $x^iy^j\in R$ with $0\le j<a$ such that $\gd(x^iy^j)=s$. Let us denote the monomial by $\phi_s$.

Let $\calP=\set{P_1,P_2,\dots,P_n}$ be a set of nonsingular affine rational points of $X$ and let $\F^n$ be the Hamming space over $\F$. The evaluation $\ev:R\to\F^n$ defined by $\phi\mapsto(\phi(P_1),\phi(P_2),\dots,\phi(P_n))$ is a linear map over $\F$. Let $u$ be a fixed positive integer less than $n$ and define
\[
	L_u=\set{f\in R\mid \gd(f)\le u}=\langle\phi_s\in R\mid s\in S, s\le u\rangle
\]
where brackets denote the linear span over $\F$. Then the AG code $C_u$ is defined as the image of $L_u$ under $\ev$. As $u<n$, the evaluation is one-to-one on $L_u$. Therefore the dimension $k$ of the code $C_u$ equals $\dim_\F L_u$, which equals the size of the set $\set{s\in S\mid s\le u}$. Let $\set{s\in S\mid s\le u}=\set{s_1,s_2,\dots,s_k}$. By nonsystematic encoding, a message 
\[
	m=(m_1,m_2,\dots,m_k)\in\F^k
\]
is encoded to the codeword $\ev(\mu)\in C_u$ where 
\[
	\mu=\sum_{i=1}^k m_i\phi_{s_i}\in L_u.
\]

For each $1\le i\le n$, let $\frakm_i=\ideal{x-\ga_i,y-\gb_i}$ be the maximal ideal of $R$ associated with the point $P_i=(\ga_i,\gb_i)$. Then we have
\[
	\frakm_i+\prod_{j\neq i}\frakm_j=\ideal{1}.
\]
Therefore there exist $g_i$ and $h_i$ such that
\[
	g_i+h_i=1,\quad g_i\in\frakm_i,\quad h_i\in \prod_{j\neq i}\frakm_j.
\]
Then $h_i(P_i)=1$ and $h_i(P_j)=0$ for $j\neq i$.  This set of $h_i$ is called a \textit{Lagrange basis} for the points $P_1,\dots,P_n$. A Lagrange basis can be easily computed as follows. Let $t$ be the number of distinct $x$-coordinates of the points $P_i$. For each of these $x$-coordinates, there are at most $a$ $y$-coordinates of the points with the same $x$-coordinate. If $h_{i,x}\in\F[x]$ and $h_{i,y}\in\F[y]$ are polynomials such that $h_{i,x}$ vanishes at the $x$-coordinates except that of $P_i$ and $h_{i,y}$ vanishes at the $y$-coordinates except that of $P_i$, then let
\[
	h_i=\frac{h_{i,x}h_{i,y}}{h_{i,x}(\ga_i)h_{i,y}(\gb_i)}\in R
\] 
Note that $\deg_x(h_{i,x}h_{i,y})=t-1$. We assume $h_i$ are precomputed prior to  decoding.

As $R$ is an $\F[x]$-module of rank $a$ with free basis $\set{y^j\mid 0\le j<a}$, a polynomial in $R[z]$ can be written as a unique $\F$-linear combination of the monomials in
\[
	\Omega=\set{x^iy^jz^k\mid 0\le i, 0\le j<a, 0\le k}.
\]
For an integer $s$, we define the weighted degee of a monomial $x^iy^jz^k\in\Omega$ by
\[
	\gd_s(x^iy^jz^k)=\gd(x^iy^j)+sk=\gd_xi+\gd_yj+sk.
\]
Using $\gd_s$, we endow a weighted degree order $>_s$ on $\Omega$, breaking ties in weighted degrees by $z>y>x$.  Note that $>_s$ restricted to the monomials belonging to $Rz\oplus R$ is a monomial order for $\F[x]$-modules. The weighted degree order restricted to $R$ is simply denoted by $>_\gd$ as it is independent of $s$. 

Note that $Rz\oplus R$ is a free $\F[x]$-module of rank $2a$ with a free basis $G=\set{y^jz,y^j\mid 0\le j<a}$. There is a simple criterion of a Gr\"obner basis of an $\F[x]$-submodule of $Rz\oplus R$ with respect to any monomial order.

\begin{prop}\label{xnskw}
Let $M$ be a submodule of $Rz\oplus R$, and let $>$ be a monomial order on $Rz\oplus R$. Suppose $B$ is a subset of $M$ that generates $M$. If elements of $B$ have leading terms that are $\F[x]$-multiples of distinct elements of $G$, then $B$ is a Gr\"obner basis of $M$ with respect to $>$. If this is the case, $B$ is also a free basis of $M$.    
\end{prop}

For a polynomial $\phi$, $\LT(\phi)$ denotes the leading term with respect to a given monomial order, and $\LC(\phi)$ denotes the coefficient of the leading term. Finally, for $f\in\F[x]$ the bracket notation $f[x^k]$ refers to the coefficient of the term $x^k$ in $f$. 

\section{Interpolation Decoding}\label{sec_kdqdd}

Let $v$ be a received vector in $\F^n$. Let $c\in C_u$ be such that $v=e+c$. Then there is a unique 
\[
	\mu=\sum_{s\in S,s\le u}\gw_s\phi_s\in L_u
\]
with $c=\ev(\mu)$. 

Let us denote the module of $z$-linear polynomials over $R$ that interpolate the points $(P_i,v_i)$ by
\[
	I_v=\set{f\in Rz\oplus R\mid f(P_i,v_i)=0, 1\le i\le n}.
\]
Then it is easy to see that $I_v=R(z-h_v)+J$ where 
\[
	h_v=\sum_{i=1}^nv_i h_i,\quad J=\bigcap_{i=1}^n\frakm_i.
\]

As $J$ is an ideal of $R$, $J$ is a free $\F[x]$-submodule of $R$ of rank $a$ and has a Gr\"obner basis $\set{\eta_0,\eta_1,\dots,\eta_{a-1}}$ with respect to $>_\gd$ such that $\deg_y(\LT(\eta_i))=i$. Then 
\begin{equation}\label{camdzs}
	\sum_{0\le i<a}\deg_x(\LT(\eta_i))=\dim_\F R/J=n.
\end{equation}
As $I_v=R(z-h_v)+J$, the set 
\begin{equation}\label{cjsjsw}
	\set{\eta_0,\eta_1,\dots,\eta_{a-1},z-h_v,y(z-h_v),\dots,y^{a-1}(z-h_v)}
\end{equation}
is a Gr\"obner basis of $I_v$ with respect to $>_{\gd(h_v)}$.

The ideal of the error vector $e$
\[
	J_e=\bigcap_{e_i\neq 0}\frakm_i
\]
also has a Gr\"obner basis $\set{\epsilon_0,\epsilon_1,\dots,\epsilon_{a-1}}$ with respect to $>_\gd$ such that $\deg_y(\LT(\epsilon_i))=i$. Then
\begin{equation}\label{mdcad}
	\sum_{0\le i<a}\deg_x(\LT(\epsilon_i))=\dim_\F R/J_e=\wt(e).
\end{equation}

The results in the following Section \ref{sec_qpwkkd} will serve as a backbone of our decoding algorithm presented in Section \ref{sec_qpwscks} and its proof in Section \ref{sec_jdmxzd}.

\subsection{Decoding by Majority Voting}\label{sec_qpwkkd}

Let $s$ be a nongap with $s\le u$. Suppose
\[
	v^{(s)}=e+\ev(\mu^{(s)}),\quad \mu^{(s)}=\gw_s\phi_s+\mu^{(s-1)},\quad \mu^{(s-1)}\in L_{s-1},
\]
and $B^{(s)}=\set{g_i^{(s)},f_i^{(s)}\mid 0\le i<a}$ is a Gr\"obner basis of $I_{v^{(s)}}$ with respect to $>_s$  where 
\[
	\begin{aligned}
	g_i^{(s)}&=\sum_{0\le j<a}c_{i,j}y^jz+\sum_{0\le j<a}d_{i,j}y^j,\quad c_{i,j},d_{i,j}\in\F[x],\\
	f_i^{(s)}&=\sum_{0\le j<a}a_{i,j}y^jz+\sum_{0\le j<a}b_{i,j}y^j,\quad a_{i,j},b_{i,j}\in\F[x]
	\end{aligned}
\]
such that $\LT(g_i^{(s)})=\LT(d_{i,i}y^i)$ and $\LT(f_i^{(s)})=\LT(a_{i,i}y^iz)$ for $0\le i<a$. Let $\nu_i^{(s)}=\LC(d_{i,i})$. 

\begin{lem}\label{lem_jjqqe}
We have
\[
	\sum_{0\le i<a}\deg(a_{i,i})+\sum_{0\le i<a}\deg(d_{i,i})=n.
\]
\end{lem}

\begin{IEEEproof}
As $B^{(s)}$ is a Gr\"obner basis of $I_{v^{(s)}}$,
\[
	\sum_{0\le i<a}\deg(a_{i,i})+\sum_{0\le i<a}\deg(d_{i,i})=\dim_{\F}(Rz\oplus R)/I_{v^{(s)}}.
\]
Since $I_{v^{(s)}}=R(z-h_{v^{(s)}})+J$, we have $\dim_{\F}(Rz\oplus R)/I_{v^{(s)}}=\dim_\F R/J=n$.
\end{IEEEproof}

\begin{lem}
For $0\le i<a$, we have $\gd(a_{i,i}y^i)\le\gd(\epsilon_i)$, equivalently $\deg(a_{i,i})\le \deg_x(\LT(\epsilon_i))$.
\end{lem}

\begin{IEEEproof}
Since $J_e(z-\mu^{(s)})\subset I_{v^{(s)}}$, we have $\epsilon_i(z-\mu^{(s)})\in I_{v^{(s)}}$. Note that $\LT(\epsilon_i(z-\mu^{(s)}))=\LT(\epsilon_iz)$ with respect to $>_s$, and $\deg_y(\epsilon_iz)=i$. As $B^{(s)}$ is a Gr\"obner basis of $I_{v^{(s)}}$, the leading term $\LT(\epsilon_iz)$ must be an $\F[x]$-multiple of $\LT(f_i^{(s)})$. Thus the assertion follows.
\end{IEEEproof}

\begin{lem}
For $0\le i<a$, we have $\gd(d_{i,i}y^{i})\le \gd(\eta_{i})$, equivalently $\deg(d_{i,i})\le \deg_x(\LT(\eta_i))$.
\end{lem}

\begin{IEEEproof}
As $B^{(s)}$ is a Gr\"obner basis of $I_{v^{(s)}}$ and $J\subset I_{v^{(s)}}$, it follows that $\eta_{i}$ is an $\F[x]$-multiple of $\LT(g_{i}^{(s)})$. Hence the assertion follows.
\end{IEEEproof}

Now let $w$ be \textit{any} element of $\F$. For each $0\le i<a$, let
\begin{equation*}
	\bar{g}_{i}=g_{i}^{(s)}(z+w\phi_s),\quad
	\bar{f}_i=f_i^{(s)}(z+w\phi_s)
\end{equation*}
where the parentheses denote substitution of the variable $z$. The automorphism of the ring $R[z]$ induced by the substitution $z\mapsto z+w\phi_s$ preserves leading terms with respect to $>_s$. Therefore the set $\bar{B}=\set{\bar{g}_i,\bar{f}_i\mid 0\le i<a}$ is a Gr\"obner basis of 
\[
	\tilde{I}=\set{f(z+w\phi_s)\mid f\in I_{v^{(s)}}}
\]
with respect to $>_s$. However, with respect to $>_{s-1}$, $\bar{B}$ is generally no longer a Gr\"obner basis of $\tilde{I}$. The following procedure modifies $\bar{B}$ to obtain a Gr\"obner basis of $\tilde{I}$ with respect to $>_{s-1}$.

For each $0\le i<a$, there are unique integers $0\le i'<a=\gd_x$ and $k_i$ satisfying
\begin{equation}\label{vnmskq}
	\gd(a_{i,i}y^i)+s=\gd_xk_i+\gd_yi'.
\end{equation}
Then let
\begin{equation}\label{jcmkwd}
	c_i=\deg_x(d_{i',i'})-k_i,\quad \bar{c}_i=\max\set{c_i,0}
\end{equation}
and
\begin{equation}\label{jhjghg}
	w_i=-\frac{b_{i,i'}[x^{k_i}]}{\mu_i},\quad 
	\mu_i=\LC(a_{i,i}y^i\phi_s).
\end{equation}
Note that the map $i\mapsto i'$ is a permutation of $\set{0,1,\dots,a-1}$ and that  the integer $c_i$ is defined such that
\begin{equation}\label{asdfer}
	\gd_xc_i=\gd(d_{i',i'}y^{i'})-\gd(a_{i,i}y^i)-s.
\end{equation}

Now if $w_i=w$, let
\begin{equation}\label{laksnc}
	\tilde{g}_{i'}=\bar{g}_{i'},\quad
	\tilde{f}_i=\bar{f}_i
\end	{equation}
and if $w_i\neq w$ and $c_i>0$, let
\begin{equation}\label{cnksls}
	\tilde{g}_{i'}=\bar{f}_i,\quad
	\tilde{f}_i=x^{c_i}\bar{f}_i-\frac{\mu_i(w-w_i)}{\nu_{i'}^{(s)}}\bar{g}_{i'}	
\end{equation}
and if $w_i\neq w$ and $c_i\le 0$, let
\begin{equation}\label{cnskjf}
	\tilde{g}_{i'}=\bar{g}_{i'},\quad
	\tilde{f}_i=\bar{f}_i-\frac{\mu_i(w-w_i)}{\nu_{i'}^{(s)}}x^{-c_i}\bar{g}_{i'}.
\end{equation}

\begin{prop}\label{cmwdq}
The set $\tilde{B}=\set{\tilde{g}_i,\tilde{f}_i\mid 0\le i<a}$ is a Gr\"obner basis of $\tilde{I}$ with respect to $>_{s-1}$.
\end{prop}

\begin{IEEEproof}
Let $0\le i<a$.  We consider the pair
\[
	\begin{aligned}
	\bar{g}_{i'}&=\sum_{0\le j<a}c_{i',j}y^jz+\sum_{0\le j<a}d_{i',j}y^j+\sum_{0\le j<a}wc_{i',j}y^j\phi_s,\\
	\bar{f}_i&=\sum_{0\le j<a}a_{i,j}y^jz+\sum_{0\le j<a}b_{i,j}y^j+\sum_{0\le j<a}wa_{i,j}y^j\phi_s.	
	\end{aligned}
\]
By the assumption that $B^{(s)}$ is a Gr\"obner basis of $I_{v^{(s)}}$ with respect to $>_s$, we have for $0\le j<a$,
\[	
	\gd(d_{i',i'}y^{i'})>\gd_s(c_{i',j}y^jz)\ge\gd(wc_{i',j}y^j\phi_s)
\]
and for $0\le j<a$ with $j\neq i'$, $\gd(d_{i',i'}y^{i'})>\gd(d_{i',j}y^j)$. Therefore with respect to $>_{s-1}$, $\LT(\bar{g}_{i'})=\LT(d_{i',i'}y^{i'})$. By the same assumption, we have for $0\le j<a$ with $j\neq i$,
\[
	\gd_s(a_{i,i}y^iz)>\gd_s(a_{i,j}y^jz)\ge\gd(wa_{i,j}y^j\phi_s)
\]
and for $0\le j<a$ with $j\neq i'$, $\gd_s(a_{i,i}y^iz)>\gd(b_{i,j}y^j)$ by the definition of $i'$ in \eqref{vnmskq}. Note that
\begin{equation}\label{ckmld}
	\gd_s(a_{i,i}y^iz)\ge\gd(b_{i,i'}y^{i'}+wa_{i,i}y^i\phi_s)
\end{equation}
where the inequality is strict if and only if $w=w_i$ by the definition of $w_i$ in \eqref{jhjghg}. 

From now on, all leading terms are with respect to $>_{s-1}$. The inequality  \eqref{ckmld} implies that if $w=w_i$, then $\LT(\bar{f}_i)=\LT(a_{i,i}y^iz)$ and if $w\neq w_i$, then $\LT(\bar{f}_i)=\LT(b_{i,i'}y^{i'}+wa_{i,i}y^i\phi_s)$.

First we consider the case that $w_i=w$. By \eqref{laksnc},
\begin{equation}\label{jfjwd}
	\LT(\tilde{g}_{i'})=\LT(\bar{g}_{i'})=\LT(d_{i',i'}y^{i'}),
	\quad
	\LT(\tilde{f}_i)=\LT(\bar{f}_i)=\LT(a_{i,i}y^iz).
\end{equation}
Next we consider the case that $w_i\neq w$ and $c_i>0$. Then we have \eqref{cnksls}. Note that
\[
	\LT(x^{c_i}\bar{f}_i)=x^{c_i}\LT(b_{i,i'}y^{i'}+wa_{i,i}y^i\phi_s),
	\quad
	\LT(\bar{g}_{i'})=\LT(d_{i',i'}y^i)
\]
and
\[
	\gd_xc_i+\gd(b_{i,i'}y^{i'}+wa_{i,i}y^i\phi_s)
	=\gd_xc_i+\gd_s(a_{i,i}y^iz)
	=\gd(d_{i',i'}y^{i'})
\]
where the second equality is from \eqref{asdfer}, and
\[
	\LC(x^{c_i}\bar{f}_i)=\LC(b_{i,i'}y^{i'}+wa_{i,i}y^i\phi_s)
	=-\mu_iw_i+\mu_iw
	=\LC(\frac{\mu_i(w-w_i)}{\nu_{i'}^{(s)}}\bar{g}_{i'}).
\]
Therefore, together with \eqref{ckmld}, 
\begin{equation}\label{ccmmdd}
	\LT(\tilde{f}_i)=\LT(x^{c_i}a_{i,i}y^iz),
	\quad
	\LT(\tilde{g}_{i'})=\LT(\bar{f}_i)=\LT(b_{i,i'}y^{i'}+wa_{i,i}y^i\phi_s).
\end{equation}
For the case that $w_i\neq w$ and $c_i\le0$, we have \eqref{cnskjf}. By repeating almost the same argument as above, we can show that
\begin{equation}\label{ckmslw}
	\LT(\tilde{g}_{i'})=\LT(d_{i',i'}y^{i'}),
	\quad
	\LT(\tilde{f}_i)=\LT(a_{i,i}y^iz).
\end{equation}
Finally it is clear that $\tilde{B}$ generates the module $\tilde{I}$. From \eqref{jfjwd}, \eqref{ccmmdd}, and \eqref{ckmslw}, we see that $\tilde{B}$ is a Gr\"obner basis of $\tilde{I}$ with respect to $>_{s-1}$, by the criterion in Proposition~\ref{xnskw}.
\end{IEEEproof}

\begin{lem}
Let $0\le i<a$. If $w_i\neq w$, then
\begin{equation}\label{zbjadd}
	\gd_{s-1}(\tilde{g}_{i'})=\gd(d_{i',i'}y^{i'})-\gd_x\bar{c}_i,		
	\quad	
	\gd_{s-1}(\tilde{f}_i)=\gd_{s-1}(a_{i,i}y^iz)+\gd_x\bar{c}_i.
\end{equation}
\end{lem}

\begin{IEEEproof}
Suppose $w_i\neq w$. Let us show the first equation. If $c_i>0$, then 
\[
	\gd_{s-1}(\tilde{g}_{i'})=\gd_{s-1}(\bar{f}_i)
	=\gd(b_{i,i'}y^{i'}+wa_{i,i}y^i\phi_s)
	=\gd_s(a_{i,i}y^iz)
	=\gd(d_{i',i'}y^{i'})-\gd_xc_i,
\]	
by \eqref{ccmmdd}, \eqref{ckmld}, and \eqref{asdfer}. If $c_i\le 0$, then $\gd_{s-1}(\tilde{g}_{i'})=\gd(d_{i',i'}y^{i'})$ by \eqref{ckmslw}. The second equation is clear by \eqref{ccmmdd} and \eqref{ckmslw}.
\end{IEEEproof}

\begin{prop}\label{jfoqff}
For $i$ with $w_i\neq \gw_s$,
\[
	\gd(\epsilon_i)-\gd(a_{i,i}y^i)\ge\gd_x\bar{c}_i
	\quad\text{and}\quad
	\min\set{\gd(\epsilon_i)+s,\gd(\eta_{i'})}\ge\gd(d_{i',i'}y^{i'}).
\]
\end{prop}

\begin{IEEEproof}
Suppose $w_i\neq\gw_s$. Then let us set $w=\gw_s$. Since $J_e(z-\gw_s\phi_s-\mu^{(s-1)})\subset I_{v^{(s)}}$, we have 
\[
	J_e(z-\mu^{(s-1)})\subset\tilde{I}.
\]
In particular, $\epsilon_i(z-\mu^{(s-1)})\in \tilde{I}$. Note that with respect to $>_{s-1}$, $\LT(\epsilon_i(z-\mu^{(s-1)}))=\LT(\epsilon_iz)$. As $\tilde{B}$ is a Gr\"obner basis of $\tilde{I}$ with respect to $>_{s-1}$, $\LT(\epsilon_iz)$ must be an $\F[x]$-multiple of the leading term of $\tilde{f}_i$. With \eqref{zbjadd}, this implies $\gd(a_{i,i}y^i)+\gd_x\bar{c}_i\le \gd(\epsilon_i)$. Now by \eqref{asdfer}, 
\[
	\gd(\epsilon_i)-\gd(a_{i,i}y^i)\ge\gd_x\bar{c}_i\ge\gd_x c_i=\gd(d_{i',i'}y^{i'})-\gd(a_{i,i}y^i)-s.
\]
Hence $\gd(\epsilon_i)+s\ge\gd(d_{i',i'}y^{i'})$.  
\end{IEEEproof}

\begin{prop}\label{kdlsef}
For $i$ with $w_i=\gw_s$,
\[
	\min\set{\gd(\epsilon_i)+s,\gd(\eta_{i'})}\ge\gd(d_{i',i'}y^{i'})-\gd_x\bar{c}_i
\]
\end{prop}

\begin{IEEEproof}
Suppose $w_i=\gw_s$. Then let us choose $w\in\F$ such that $w\neq w_i$. Since $J_e(z-\gw_s\phi_s-\mu^{(s-1)})\subset I_{v^{(s)}}$, we have 
\[
	J_e(z-(\gw_s-w)\phi_s-\mu^{(s-1)})\subset \tilde{I}.
\]
In particular, $\epsilon_i(z-(\gw_s-w)\phi_s-\mu^{(s-1)})\in \tilde{I}$. Note that $\gw_s-w\neq 0$. With respect to $>_{s-1}$, 
\[
	\LT(\epsilon_i(z-(\gw_s-w)\phi_s-\mu^{(s-1)}))=\LT((\gw_s-w)\epsilon_i\phi_s)
\]
As $\tilde{B}$ is a Gr\"obner basis of $\tilde{I}$ with respect to $>_{s-1}$, $\LT((\gw_s-w)\epsilon_i\phi_s)$ must be a scalar multiple of the leading term of $\tilde{g}_{i'}$. With \eqref{zbjadd}, this implies $\gd(\epsilon_i)+s\ge\gd(d_{i',i'}y^{i'})-\gd_x\bar{c}_i$. Finally, $\gd(\eta_{i'})\ge \gd(d_{i',i'}y^{i'})\ge \gd(d_{i',i'}y^{i'})-\gd_x\bar{c}_i$.
\end{IEEEproof}

\begin{prop}
The condition
\[
	\sum_{0\le i<a}\max\set{\gd(\eta_{i'})-\gd(y^i)-s,\gd(\epsilon_i)-\gd(y^i)}>2\gd_x\wt(e)
\]
implies
\[
	\sum_{w_i=\gw_s}\bar{c}_i>\sum_{w_i\neq\gw_s}\bar{c}_i.
\]
\end{prop}

\begin{IEEEproof}
Propositions \ref{jfoqff} and \ref{kdlsef} imply
\[
	\begin{aligned}
	\sum_{w_i=\gw_s}\gd_x\bar{c}_i
	&\ge\sum_{w_i=\gw_s}\gd(d_{i',i'}y^{i'})-\min\set{\gd(\epsilon_i)+s,\gd(\eta_{i'})}\\
	&\ge\sum_{0\le i<a}\gd(d_{i',i'}y^{i'})-\min\set{\gd(\epsilon_i)+s,\gd(\eta_{i'})}
	\end{aligned}
\]
and
\[
	\sum_{w_i\neq\gw_s}\gd_x\bar{c}_i
	\le\sum_{w_i\neq\gw_s}\gd(\epsilon_i)-\gd(a_{i,i}y^i)
	\le\sum_{0\le i<a}\gd(\epsilon_i)-\gd(a_{i,i}y^i).
\]
Now we have a chain of equivalent conditions
\[
	\begin{aligned}
	&\sum_{0\le i<a}\gd(d_{i',i'}y^{i'})-\min\set{\gd(\epsilon_i)+s,\gd(\eta_{i'})}
	>\sum_{0\le i<a}\gd(\epsilon_i)-\gd(a_{i,i}y^i)\\
	&\iff \sum_{0\le i<a}\gd(d_{i',i'}y^{i'})+\sum_{0\le i<a}\gd(a_{i,i}y^i)-\min\set{\gd(\epsilon_i)+s,\gd(\eta_{i'})}
	>\sum_{0\le i<a}\gd(\epsilon_i)\\
	&\iff \sum_{0\le i<a}\gd(\eta_{i'})+\sum_{0\le i<a}\gd(y^i)+\max\set{-\gd(\epsilon_i)-s,-\gd(\eta_{i'})}
	>\sum_{0\le i<a}\gd(\epsilon_i)\\	
	&\iff \sum_{0\le i<a}\max\set{\gd(\eta_{i'})-\gd(y^i)-s,\gd(\epsilon_i)-\gd(y^i)}
	>\sum_{0\le i<a}2(\gd(\epsilon_i)-\gd(y^i))				
	\end{aligned}
\]
where we used the equality
\[
	\begin{split}
 	\sum_{0\le i<a}\gd(d_{i',i'}y^{i'})+\sum_{0\le i<a}\gd(a_{i,i}y^i)
	&=\sum_{0\le i<a}\gd(d_{i,i}y^i)+\sum_{0\le i<a}\gd(a_{i,i}y^i)\\
	&=\sum_{0\le i<a}(\gd(d_{i,i})+\gd(a_{i,i}))+\sum_{0\le i<a}2\gd(y^i)\\
	&=\gd_xn+\sum_{0\le i<a}2\gd(y^i)\\
	&=\sum_{0\le i<a}(\gd(\eta_{i})-\gd(y^i))+\sum_{0\le i<a}2\gd(y^i)\\
	&=\sum_{0\le i<a}\gd(\eta_{i'})+\sum_{0\le i<a}\gd(y^i)
	\end{split}
\]
shown by Lemma \ref{lem_jjqqe} and \eqref{camdzs}. Finally note that
\[
	\sum_{0\le i<a}2(\gd(\epsilon_i)-\gd(y^i))=\sum_{0\le i<a}2\gd_x\deg_x(\epsilon_i)=2\gd_x\wt(e)
\]
by \eqref{mdcad}.
\end{IEEEproof}

\begin{prop}\label{qwerrd}
Let
\[
	\nu(s)=\frac{1}{\gd_x}\sum_{0\le i<a}\max\set{\gd(\eta_{i'})-\gd(y^i)-s,0}.
\]
The condition $\nu(s)>2\wt(e)$ implies
\[
	\sum_{w_i=\gw_s}\bar{c}_i>\sum_{w_i\neq\gw_s}\bar{c}_i.
\]
\end{prop}

\begin{IEEEproof}
We have
\[
	\sum_{0\le i<a}\max\set{\gd(\eta_{i'})-\gd(y^i)-s,\gd(\epsilon_i)-\gd(y^i)}
	\ge\sum_{0\le i<a}\max\set{\gd(\eta_{i'})-\gd(y^i)-s,0}
\]
as $\gd(\epsilon_i)-\gd(y^i)\ge 0$ for $0\le i<a$.
\end{IEEEproof}

\subsection{Decoding Algorithm}\label{sec_qpwscks}

\paragraph{Initialization} Let $N=\gd(h_v)$, and let $B^{(N)}$ be the Gr\"obner basis of $I_v$ with respect to $>_N$ given in \eqref{cjsjsw}. The steps \textit{Pairing, Voting, Rebasing} are iterated for $s$ decreasing from $N$ to $0$.

\paragraph{Pairing} Suppose $B^{(s)}=\set{g_i^{(s)},f_i^{(s)}\mid 0\le i<a}$ is a Gr\"obner basis of $I_{v^{(s)}}$ with respect to $>_s$ where 
\[
	\begin{aligned}
	g_i^{(s)}&=\sum_{0\le j<a}c_{i,j}y^jz+\sum_{0\le j<a}d_{i,j}y^j\\
	f_i^{(s)}&=\sum_{0\le j<a}a_{i,j}y^jz+\sum_{0\le j<a}b_{i,j}y^j
	\end{aligned}
\]
and let $\nu_i^{(s)}=\LC(d_{i,i})$. For $0\le i<a$, there are unique integers $0\le i'<\gd_x=a$ and $k_i$ satisfying
\[
	\gd(a_{i,i}y^i)+s=\gd_xk_i+\gd_yi'.
\]
Note that the integer $\gd(a_{i,i}y^i)+s$ is a nongap if and only if $k_i\ge 0$. Now let
\[
	c_i=\deg_x(d_{i',i'})-k_i.
\]

\paragraph{Voting}
If $s>u$ or $s$ is a gap, then for $i$ with nongap $\gd(a_{i,i}y^i)+s$, let
\[
	w_i=-b_{i,i'}[x^{k_i}],\quad \mu_i=1
\]
and for $i$ with gap $\gd(a_{i,i}y^i)+s$, let $w_i=0, \mu_i=1$. Let $w=0$ in both cases.

If  $s\le u$ and $s$ is a nongap, then for each $i$, we let
\[
	w_i=-\frac{b_{i,i'}[x^{k_i}]}{\mu_i},\quad \mu_i=\LC(a_{i,i}y^i\phi_s)
\]
and let $\bar{c}_i=\max\set{c_i,0}$, and let $w$ be the element of $\F$ with the largest
\[
	\sum_{w=w_i}\bar{c}_i,
\]
and let $w_s=w$.

\paragraph{Rebasing} For each $i$, we do the following. If $w_i=w$, then let
\begin{equation}\label{fkfmvd}
	\begin{aligned}
	g_{i'}^{(s-1)}&=g_{i'}^{(s)}(z+w\phi_s)\\
	f_i^{(s-1)}&=f_i^{(s)}(z+w\phi_s)
	\end{aligned}
\end{equation}
and let $\nu_{i'}^{(s-1)}=\nu_{i'}^{(s)}$. 
If $w_i\neq w$ and $c_i>0$, then let
\begin{equation}\label{fkmksd}
	\begin{aligned}
	g_{i'}^{(s-1)}&=f_i^{(s)}(z+w\phi_s)\\
	f_i^{(s-1)}&=x^{c_i}f_i^{(s)}(z+w\phi_s)-\frac{\mu_i(w-w_i)}{\nu_{i'}^{(s)}}g_{i'}^{(s)}(z+w\phi_s)
	\end{aligned}	
\end{equation}
and let $\nu_{i'}^{(s-1)}=\mu_i(w-w_i)$.
If $w_i\neq w$ and $c_i\le 0$, then let
\begin{equation}\label{akfjcd}
	\begin{aligned}
	g_{i'}^{(s-1)}&=g_{i'}^{(s)}(z+w\phi_s)\\
	f_i^{(s-1)}&=f_i^{(s)}(z+w\phi_s)-\frac{\mu_i(w-w_i)}{\nu_{i'}^{(s)}}x^{-c_i}g_{i'}^{(s)}(z+w\phi_s)
	\end{aligned}
\end{equation}
and let $\nu_{i'}^{(s-1)}=\nu_{i'}^{(s)}$. Let $B^{(s-1)}=\set{g_i^{(s-1)},f_i^{(s-1)}\mid 0\le i<a}$.

\paragraph{Termination}
After the iterations, output the recovered message $(w_{s_1},w_{s_2},\dots,w_{s_k})$.

\subsection{Proof of the Algorithm}\label{sec_jdmxzd}

Let us start with a brief overview of the algorithm. Note that the decoding algorithm is in one of two phases while $s$ decreases from $N$ to $0$. The first phase is when $s>u$ or $s$ is a gap, and the second phase is when $s\le u$ and $s$ is a nongap. Let $v^{(N)}=v$. In the first phase,  the Gr\"obner basis $B^{(s)}$ of $I_{v^{(s)}}$ with respect to $>_s$ is updated such that $B^{(s-1)}$ is a Gr\"obner basis of $I_{v^{(s-1)}}$ with respect to $>_{s-1}$ where
\[
	v^{(s-1)}=v^{(s)}.
\]
In the second phase, the algorithm determines $w_s$ by majority voting and updates $B^{(s)}$ such that $B^{(s-1)}$ is a Gr\"obner basis of $I_{v^{(s-1)}}$ with respect to $>_{s-1}$ where  
\[
	v^{(s-1)}=v^{(s)}-\ev(w_s\phi_s).
\]
When the algorithm terminates, $w_s$ are determined for all nongaps $s\le u$.

\begin{prop}\label{cmskqe}
For $N\ge s\ge 0$, the set $B^{(s)}$ is a Gr\"obner basis of $I_{v^{(s)}}$ with respect to $>_s$.
\end{prop}

\begin{IEEEproof}
This is proved by induction on $s$. For $s=N$, this is true by \eqref{cjsjsw}. Now our induction assumption is that this is true for $s$. In the second phase, we already saw in Proposition \ref{cmwdq} that $B^{(s-1)}$ is a Gr\"obner basis of $I_{v^{(s-1)}}$. So it remains to consider the first phase. The proof for this case is similar to that of Proposition \ref{cmwdq}.

Suppose $s>u$ or $s$ is a gap. Let $0\le i<a$. Recall
\[
	\begin{aligned}
	g_{i'}^{(s)}&=\sum_{0\le j<a}c_{i',j}y^jz+\sum_{0\le j<a}d_{i',j}y^j\\
	f_i^{(s)}&=\sum_{0\le j<a}a_{i,j}y^jz+\sum_{0\le j<a}b_{i,j}y^j
	\end{aligned}
\]
By the induction assumption, we have for $0\le j<a$,
\[	
	\gd(d_{i',i'}y^{i'})>\gd_s(c_{i',j}y^jz)=\gd(c_{i',j}y^j)+s
\]
and for $0\le j<a$ with $j\neq i'$, $\gd(d_{i',i'}y^{i'})>\gd(d_{i',j}y^j)$. Therefore with respect to $>_{s-1}$, $\LT(g_{i'}^{(s)})=\LT(d_{i',i'}y^{i'})$.
Similarly, by the induction assumption, we have for $0\le j<a$ with $j\neq i$, $\gd_s(a_{i,i}y^iz)>\gd_s(a_{i,j}y^jz)$ and for $0\le j<a$ with $j\neq i'$, $\gd_s(a_{i,i}y^iz)>\gd(b_{i,j}y^j)$.

Note that
\begin{equation}\label{njlkih}
	\gd_s(a_{i,i}y^iz)\ge\gd(b_{i,i'}y^{i'})
\end{equation}
where the inequality is strict except when $\gd(a_{i,i}y^i)+s$ is a nongap and $b_{i,i'}[x^{k_i}]\neq 0$. 
Note that $w_i=0$ if and only if $\gd(a_{i,i}y^i)+s$ is a gap or $\gd(a_{i,i}y^i)+s$ is a nongap but $b_{i,i'}[x^{k_i}]=0$. Therefore with respect to $>_{s-1}$ if $w_i=0$, then $\LT(f_i^{(s)})=\LT(a_{i,i}y^iz)$ and if $w_i\neq 0$, then $\LT(f_i^{(s)})=\LT(b_{i,i'}y^{i'})$.

Now let us consider the case when $w_i=0$, then by \eqref{fkfmvd} and \eqref{njlkih},
\[
	\LT(g_{i'}^{(s-1)})=\LT(g_{i'}^{(s)})=\LT(d_{i',i'}y^{i'}),
	\quad
	\LT(f_i^{(s-1)})=\LT(f_i^{(s)})=\LT(a_{i,i}y^iz).
\]
with respect to $>_{s-1}$.
We consider the case when $w_i\neq 0$ and $c_i>0$. Then by \eqref{fkmksd},
\[
	g_{i'}^{(s-1)}=f_i^{(s)},
	\quad
	f_i^{(s-1)}=x^{c_i}f_i^{(s)}+\frac{\mu_iw_i}{\nu_{i'}^{(s)}}g_{i'}^{(s)}.
\]
Note that
\[
	\begin{gathered}
	\LT(x^{c_i}f_i^{(s)})=x^{c_i}\LT(b_{i,i'}y^{i'}),
	\quad
	\LT(g_{i'}^{(s)})=\LT(d_{i',i'}y^i),\\
	c_i\gd_x+\gd(b_{i,i'}y^{i'})
	=c_i\gd_x+\gd(a_{i,i}y^iz)
	=\gd(d_{i',i'}y^{i'}),
	\end{gathered}
\]
and
\[
	\LC(x^{c_i}f_i^{(s)})=\LC(b_{i,i'}y^{i'})=-\mu_iw_i
	=-\LC(\frac{\mu_iw_i}{\nu_{i'}^{(s)}}g_{i'}^{(s)}).
\]
Together with \eqref{njlkih}, this implies $\LT(f_i^{(s-1)})=\LT(x^{c_i}a_{i,i}y^iz)$.

For the case when $w_i\neq 0$ and $c_i\le0$, we have by \eqref{akfjcd}
\[
	g_{i'}^{(s-1)}=g_{i'}^{(s)},
	\quad
	f_i^{(s-1)}=f_i^{(s)}+\frac{\mu_iw_i}{\nu_{i'}^{(s)}}x^{-c_i}g_{i'}^{(s)}.
\]
By the same argument as when $c_i>0$, we can show that $\LT(f_i^{(s-1)})=\LT(a_{i,i}y^iz)$.

Hence the set $B^{(s-1)}$ is again a Gr\"obner basis of $I_{v^{(s-1)}}$ with respect to $>_{s-1}$.
\end{IEEEproof}

\begin{prop}
Let 	
\[
	d_u=\min\set{\nu(s)\mid s\in S,s\le u}.
\]
Then $d_u\ge n-u$. If $2\wt(e)<d_u$, then $w_s=\gw_s$ for all $s\in S,s\le u$. Hence
\[
	\sum_{s\in S,s\le u} w_s\phi_s=\mu.
\]
\end{prop}

\begin{IEEEproof}
The bound $d_u\ge n-u$ follows from
\[
	\begin{split}
	\nu(s)&=\frac{1}{\gd_x}\sum_{0\le i<a}\max\set{\gd(\eta_{i'})-\gd(y^i)-s,0}\\
	&\ge\frac{1}{\gd_x}\sum_{0\le i<a}(\gd(\eta_{i'})-\gd(y^i)-s)\\
	&=\frac{1}{\gd_x}\sum_{0\le i<a}(\gd(\eta_{i})-\gd(y^i))-s=n-s.
	\end{split}
\]
If we suppose  $2\wt(e)<d_u$, then Propositions \ref{qwerrd} and \ref{cmskqe} imply $w_s=\gw_s$ for all $s\in S,s\le u$. 
\end{IEEEproof}

\section{Hermitian Codes}\label{sec_dkdwas}

A Hermitian curve $H$ is a smooth plane curve defined with the equation $y^q+y=x^{q+1}$ over $\F_{q^2}$. It has $q^3$ rational points $P_i$ with a unique nonsingular point $P_\infty$ at infinity. The functions $x$ and $y$ on $H$ have poles at $P_\infty$ of orders $q$ and $q+1$, respectively. That is, $\gd_x=q$, $\gd_y=q+1$. 

The ideal $J$ associated with the sum of $P_i$ is an $\F[x]$-module generated by
\[
	\eta_i=y^i(x^{q^2}-x),\quad 0\le i<q
\]
which form a Gr\"obner basis of $J$ with respect to $>_\gd$.

\begin{prop}
For nongap $s<q^3$,
\[
	\nu(s)=(q-r)(q^2+r-t)+r\max\set{q^2+r-q-t-1,0}
\]
where  $s=tq+r$, $0\le r<q$.
\end{prop}

\begin{IEEEproof}
Suppose
\[
	\gd(y^i)+s=(q+1)i+s=aq+i'(q+1),\quad 0\le i'<q.
\]
Then $\gd(\eta_{i'})=q^3+i'(q+1)$. As $i'=(s+i)\bmod q$, 
\[
	\begin{aligned}
	\nu(s)&=\frac{1}{q}\sum_{i=0}^{q-1}\max\set{\gd(\eta_{i'})-\gd(y^i)-s,0}\\
	&=\frac{1}{q}\sum_{i=0}^{q-1}\max\set{q^3+((s+i)\bmod q)(q+1)-(q+1)i-s,0}\\
	&=\frac{1}{q}\sum_{i=0}^{q-1}\max\set{q^3-qi+((s+i)\bmod q)q-((s+i)-(s+i)\bmod q),0}\\
	&=\sum_{i=0}^{q-1}\max\set{q^2-i+(s+i)\bmod q-\lfloor{(s+i)/q}\rfloor,0}.
	\end{aligned}
\]
Now let $s=tq+r$, $0\le t<q^2$, $0\le r<q$. Then
\[
	\begin{aligned}
	\nu(s)&=\sum_{i=0}^{q-1}\max\set{q^2-i+(r+i)\bmod q-t-\lfloor{(r+i)/q}\rfloor,0}\\
	&=\sum_{i=0}^{q-1-r}\max\set{q^2-i+r+i-t,0}
	+\sum_{i=q-r}^{q-1}\max\set{q^2-i+r+i-q-t-1,0}\\
	&=(q-r)(q^2+r-t)+r\max\set{q^2+r-q-t-1,0}.\\
	\end{aligned}
\]
\end{IEEEproof}

\begin{prop}
For nongap $u<q^3$,
\[
	\begin{aligned}
	d_u=\min\set{\nu(s)\mid \text{nongap $s\le u$}}
	=\begin{cases}
	q^3-aq & \text{if $b \le a-(q^2-q)$},\\
	q^3-u & \text{if $b > a-(q^2-q)$},
	\end{cases}
	\end{aligned}
\]
where $u=aq+b$, $0\le b<q$.
\end{prop}

\begin{IEEEproof}
For nongap $s=tq+r$,
\[
	\begin{aligned}
	\nu(s)&=(q-r)(q^2+r-t)+r\max\set{q^2+r-q-t-1,0}\\
	&=q^3-tq+r\max\set{-1,t-q^2+q-r}\ge q^3-s.
	\end{aligned}
\]
So we see that if $a-q^2+q-b\ge 0$, that is, $b\le a-(q^2-q)$, then the minimum
\[
	\nu(aq)= q^3-aq
\]
is attained when $s=aq$, and hence $d_u=q^3-aq$. On the other hand, if $b>a-(q^2-q)$, then 
\[
	\nu(u)=q^3-aq-b=q^3-u
\]
is the minimum, and therefore $d_u=q^3-u$.
\end{IEEEproof}

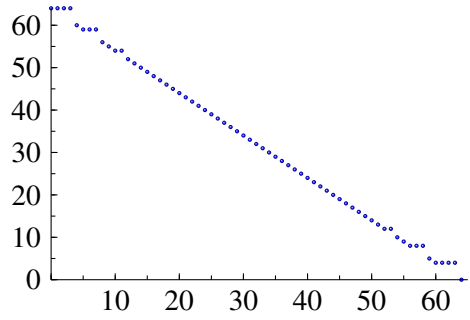
\begin{figure}[htbp]
\begin{center}
\begin{tikzpicture}[scale=.8,x=3pt,y=2pt,
	declare function={
		q = 4;
		b(\u) = mod(\u,q);
		a(\u) = div(\u,q);
		d(\u) = b(\u) <= a(\u) - (q^2 - q) ? q^3 - a(\u)*q : q^3 - \u;
		s(\u)= a(\u) - b(\u) < 0 ? a(\u)*q+a(\u) : \u;
		del(\u)= d(s(\u)); 
	}
]
	\draw[samples at={0,1,...,64},mark=ball,mark size=1pt,only marks,variable=\u] 
		plot (\u,{del(\u)});	
	\draw (0,0) -- (65,0);
	\draw (0,0) -- (0,64);
	\foreach \x in {0,10,...,64}
		\draw (\x,0) -- (\x,3pt);	
	\foreach \x in {0,5,...,64}
		\draw (\x,0) -- (\x,2pt);		
	\foreach \x in {10,20,...,64}
		\draw (\x,0) node[anchor=north] {\x};	
	\foreach \y in {0,10,...,64}
		\draw (0,\y) -- (3pt,\y);
	\foreach \y in {0,5,...,64}
		\draw (0,\y) -- (2pt,\y);
	\foreach \y in {0,10,...,64}
		\draw (0,\y) node[anchor=east] {\y};			
\end{tikzpicture}
\caption{Order bound for Hermitian code over $\F_{16}$}
\label{fig1}
\end{center}
\end{figure}

\begin{figure}[htbp]
\begin{center}
\tikzset{external/remake next}
\begin{tikzpicture}
\begin{scope}[scale=.8,x=.6pt,y=.4pt,
	declare function={
		q = 8;
		b(\u) = mod(\u,q);
		a(\u) = div(\u,q);
		d(\u) = b(\u) <= a(\u) - (q^2 - q) ? q^3 - a(\u)*q : q^3 - \u;
		s(\u)= a(\u) - b(\u) < 0 ? a(\u)*q+a(\u) : \u;
		del(\u)= d(s(\u)); 
	}
]
	\draw[samples at={0,1,...,512},mark=ball,mark size=.3pt,only marks,variable=\u] 
		plot (\u,{del(\u)});	
	\draw (0,0) -- (512,0);
	\draw (0,0) -- (0,512);
	\foreach \x in {0,100,...,512}
		\draw (\x,0) -- (\x,3pt);	
	\foreach \x in {0,50,...,512}
		\draw (\x,0) -- (\x,2pt);		
	\foreach \x in {50,100,...,512}
		\draw (\x,0) node[anchor=north] {\x};
	\foreach \y in {0,100,...,512}
		\draw (0,\y) -- (3pt,\y);
	\foreach \y in {0,50,...,512}
		\draw (0,\y) -- (2pt,\y);
	\foreach \y in {0,50,...,512}
		\draw (0,\y) node[anchor=east] {\y};	
	\draw[red] (445,-3) -| (515,67) -| (445,-3);	
\end{scope}			
\begin{scope}[scale=.8,x=3pt,y=2pt,shift={(-350,30)},
	declare function={
		q = 8;
		b(\u) = mod(\u,q);
		a(\u) = div(\u,q);
		d(\u) = b(\u) <= a(\u) - (q^2 - q) ? q^3 - a(\u)*q : q^3 - \u;
		s(\u)= a(\u) - b(\u) < 0 ? a(\u)*q+a(\u) : \u;
		del(\u)= d(s(\u)); 
	}
]
	\draw[samples at={448,449,...,512},mark=ball,mark size=1pt,only marks,variable=\u] 
		plot (\u,{del(\u)});	
	\draw (448,0) -- (512,0);
	\draw (445,0) -- (445,65);
	\foreach \x in {450,460,...,512}
		\draw (\x,0) -- (\x,3pt);	
	\foreach \x in {450,455,...,512}
		\draw (\x,0) -- (\x,2pt);		
	\foreach \x in {450,460,...,512}
		\draw (\x,0) node[anchor=north] {\x};
	\foreach \y in {0,10,...,65}
		\draw (445,0) +(0,\y) -- +(3pt,\y);
	\foreach \y in {0,5,...,65}
		\draw (445,0) +(0,\y) -- +(2pt,\y);
	\foreach \y in {0,10,...,65}
		\draw (445,0) +(0,\y) node[anchor=east] {\y};
	\draw[red] (447,-1) -| (513,65) -| (447,-1);	
\end{scope}								
\end{tikzpicture}
\caption{Order bound for Hermitian codes over $\F_{64}$}
\label{fig2}
\end{center}
\end{figure}
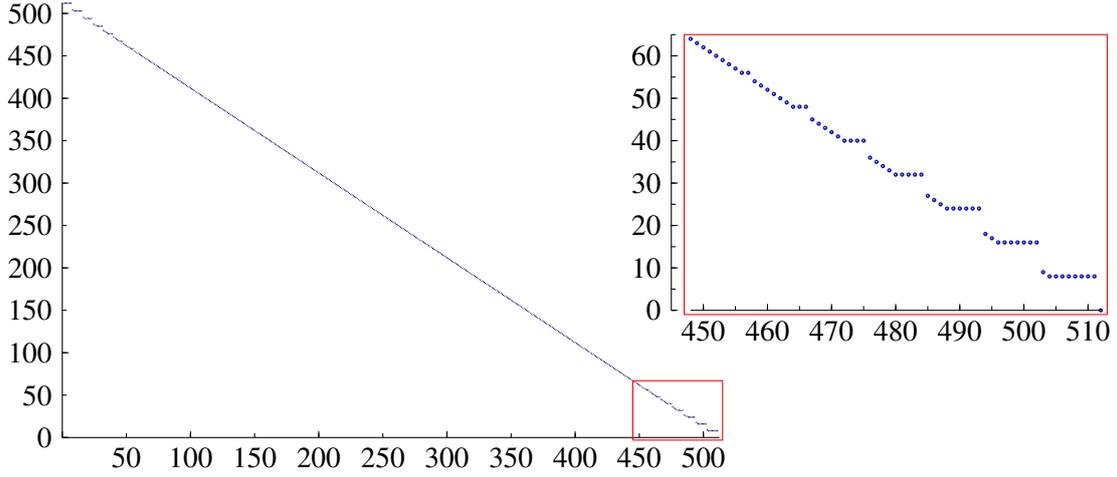

It can be shown that the bound $d_u$ exactly matches with the order bound on the minimum distance of Hermitian codes as given in \cite{maria2007} and \cite{hoholdt1998}. Hence we may call $d_u$ also an order bound. Figures \ref{fig1} and \ref{fig2} show the order bounds for Hermitian codes with $q=3$ and $q=8$, respectively.

Let $\F_{9}=\F_{3}(\ga)$ with $\ga^2-\ga-1=0$. We use the Hermitian curve over $\F_{9}$ defined by $y^3+y=x^4$, which has $27$ rational points 
\[
	\begin{gathered}
	( 0, 0 ), ( 0, \ga^2 ), ( 0, \ga^6 ), ( 1, 2 ), ( 1, \ga ), ( 1, \ga^3 ), ( 2, 2 ), ( 2, \ga ), ( 2, \ga^3 ),
 	( \ga, 1 ), ( \ga, \ga^7 ), ( \ga, \ga^5 ), ( \ga^2, 2 ), ( \ga^2, \ga ), \\
	( \ga^2, \ga^3 ), ( \ga^7, 1 ), ( \ga^7, \ga^7 ), ( \ga^7, \ga^5 ),
	 ( \ga^5, 1 ), ( \ga^5, \ga^7 ), ( \ga^5, \ga^5 ), ( \ga^3, 1 ), ( \ga^3, \ga^7 ), ( \ga^3, \ga^5 ), ( \ga^6, 2 ), ( \ga^6, \ga ), ( \ga^6, \ga^3 )
	\end{gathered}
\]
to define the Hermitian code $C_{16}$, $[27,14,11]$ linear code over $\F_{9}$.

Suppose that the sent codeword was corrupted during the transmission, and the received vector is
\[
	v=(0,0,0,0,0,\ga^2,2,0,0,0,0,0,0,0,0,0,0,0,0,\ga^3,0,0,\ga^7,0,0,2,0).
\]

The six generators of the module $I_v$ are
\[
	\begin{aligned}
	g_0&=x^9-x,\\
	g_1&=y(x^9-x),\\
	g_2&=y^2(x^9-x),\\
	f_0&=z-h_v,\\
	f_1&=y(z-h_v),\\
	f_2&=y^2(z-h_v),
	\end{aligned}
\]
where
\[
	\begin{split}
	h_v&=(\ga^3x^8 + x^7 + \ga^7x^6 + \ga^5x^5 + \ga^3x^4 + \ga^6x^3 + 2x^2 + \ga^2x)y^2\\
	&\quad + (\ga^6x^8 + x^7 + x^6 + ax^5 + \ga^6x^4 + \ga^7x^3 + \ga^3x)y \\
	&\quad + 2x^8 + ax^7 + \ga^6x^6 + 2x^4 + \ga^2x^3 + x.
	\end{split}
\]
Note that $N=\gd(h_v)=32$. Thus the initial basis for the code $C_{16}$ is
\[
\begin{array}{r*{6}{|r}}
 &\makebox[50pt][r]{$y^2z$}&\makebox[50pt][r]{$yz$}&\makebox[50pt][r]{$z$}
 &\makebox[50pt][r]{$y^2$}&\makebox[50pt][r]{$y$}&\makebox[50pt][r]{$1$}\\
\hline
g_0& &  &  &  &  & x^9+\cdots \\
g_1& &  &  &  &x^9+\cdots& \\
g_2& &  &  &x^9+\cdots&  & \\
f_0& &  & 1 &\ga^7x^8+\cdots&\ga^2x^8+\cdots & x^8+\cdots \\
f_1& & 1&  & \ga^2x^8+\cdots &\ga^6x^8+\cdots&\ga^7x^{12}+\cdots\\
f_2& 1&  &  &\ga^6x^8+\cdots &\ga^7x^{12}+\cdots& \ga^2x^{12}+\cdots
\end{array}
\]
which is a Gr\"obner basis with respect to $>_{32}$. In \textit{Pairing} and \textit{Voting} steps, the following data is computed:
\[
\begin{array}{ccrr}
\hline
 f_i&g_{i'}&c_i&w_i \\
 \hline
f_0&g_2&1&\ga^3\\
f_1&g_0&-3&\ga^3\\
f_2&g_1&-3&\ga^3
\end{array}
\]
In \textit{Rebasing} step, the pair $f_0,g_2$ is modified by \eqref{fkmksd} while the pairs $f_1,g_0$ and $f_2,g_1$ are modified by \eqref{akfjcd}. These modifications give the Gr\"obner basis with respect to $>_{31}$,
\[
\begin{array}{r*{6}{|r}}
 &\makebox[50pt][r]{$y^2z$}&\makebox[50pt][r]{$yz$}&\makebox[50pt][r]{$z$}
 &\makebox[50pt][r]{$y^2$}&\makebox[50pt][r]{$y$}&\makebox[50pt][r]{$1$}\\
\hline
g_0& &  &  &  &  & x^9+\cdots \\
g_1& &  &  &  &x^9+\cdots& \\
g_2& &  &  &\ga^7x^8+\cdots&\ga^2x^8+\cdots  &x^8+\cdots \\
f_0& &  & x &2x^8+\cdots&\ga^2x^9+\cdots & x^9+\cdots \\
f_1& & 1&  & \ga^2x^8+\cdots &\ga^6x^8+\cdots&2x^{11}+\cdots\\
f_2& 1&  &  &\ga^6x^8+\cdots &2x^{11}+\cdots& \ga^2x^{12}+\cdots
\end{array}
\]
After similar iterations, we eventually reach to the Gr\"obner basis with respect to $>_{16}$ for $v$,
\[
\begin{array}{r*{6}{|r}}
 &\makebox[50pt][r]{$y^2z$}&\makebox[50pt][r]{$yz$}&\makebox[50pt][r]{$z$}
 &\makebox[50pt][r]{$y^2$}&\makebox[50pt][r]{$y$}&\makebox[50pt][r]{$1$}\\
\hline
g_0 & & & & & &x^9+\cdots \\
g_1 & &x+\cdots&\ga^7x^2+\cdots&2x^5+\cdots&x^7+\cdots&\ga^7x^8+\cdots \\
g_2 & & &x+\cdots&\ga^2x^7+\cdots&\ga^5x^8+\cdots&x^9+\cdots \\
f_0 & &1&x^2+\cdots& & &\\
f_1 & &x^2+\cdots&\ga^7x^3+\cdots&\ga^7x^4+\cdots&\ga^3x^6+\cdots&\ga^7x^8+\cdots \\
f_2 & 1&\ga^3x+\cdots&\ga^7x^2+\cdots&\ga^3x^4+\cdots&\ga^7x^6+\cdots&\ga^3x^8+\cdots
\end{array}
\]
In \textit{Pairing} and \textit{Voting} steps, the following data is computed:
\[
\begin{array}{ccrr}
\hline
f_i&g_{i'}&c_i&w_i \\
\hline
f_0&g_1&1&0\\
f_1&g_2&1&0\\
f_2&g_0&1&\ga^7
\end{array}
\]
Thus the value $0$ gets $2$ votes, and the value $\ga^7$ gets $1$ vote. So $w$ is set to be $0$, and this result is recorded in $w_{16}=0$. Then the pairs $f_0$, $g_1$ and $f_1$, $g_2$ are modified by \eqref{fkfmvd}. The pair $f_2$, $g_0$ is modified by \eqref{fkmksd}. Then we get the Gr\"obner basis with respect to $>_{15}$ for $v^{(15)}=v-\ev(0\cdot x^4y)$,
\[
\begin{array}{r*{6}{|r}}
 &\makebox[50pt][r]{$y^2z$}&\makebox[50pt][r]{$yz$}&\makebox[50pt][r]{$z$}
 &\makebox[50pt][r]{$y^2$}&\makebox[50pt][r]{$y$}&\makebox[50pt][r]{$1$}\\
\hline
g_0 &1&\ga^3x+\cdots&\ga^7x^2+\cdots&\ga^3x^4+\cdots&\ga^7x^6+\cdots&\ga^3x^8+\cdots \\
g_1 & &x+\cdots&\ga^7x^2+\cdots&2x^5+\cdots&x^7+\cdots&\ga^7x^8+\cdots \\
g_2 & & &x+\cdots&\ga^2x^7+\cdots&\ga^5x^8+\cdots&x^9+\cdots \\
f_0 & &1&x^2+\cdots& & &\\
f_1 & &x^2+\cdots&\ga^7x^3+\cdots&\ga^7x^4+\cdots&\ga^3x^6+\cdots&\ga^7x^8+\cdots \\
f_2 &x+\cdots&\ga^3x^2+\cdots&\ga^7x^3+\cdots&\ga^3x^5+\cdots&\ga^7x^7+\cdots&\ga^5x^8+\cdots
\end{array}
\]
In \textit{Pairing} and \textit{Voting} steps, we obtain
\[
\begin{array}{ccrr}
\hline
f_i&g_{i'}&c_i&w_i \\
\hline
f_0&g_0&1&0\\
f_1&g_1&0&0\\
f_2&g_2&1&0
\end{array}
\]
So $w_{15}=w=0$. All three pairs $f_0,g_0$, $f_1,g_1$, and $f_2,g_2$ are modified by \eqref{fkfmvd}.
Thus we get the Gr\"obner basis with respect to $>_{14}$ for $v^{(14)}=v^{(15)}-\ev(0\cdot x^5)$,
\[
\begin{array}{r*{6}{|r}}
 &\makebox[50pt][r]{$y^2z$}&\makebox[50pt][r]{$yz$}&\makebox[50pt][r]{$z$}
 &\makebox[50pt][r]{$y^2$}&\makebox[50pt][r]{$y$}&\makebox[50pt][r]{$1$}\\
\hline
g_0&1&\ga^3x+\cdots&\ga^7x^2+\cdots&\ga^3x^4+\cdots&\ga^7x^6+\cdots&\ga^3x^8+\cdots \\
g_1& &x+\cdots&\ga^7x^2+\cdots&2x^5+\cdots&x^7+\cdots&\ga^7x^8+\cdots \\
g_2& & &x+\cdots&\ga^2x^7+\cdots&\ga^5x^8+\cdots&x^9+\cdots \\
f_0& &1&x^2+\cdots& & &\\
f_1& &x^2+\cdots&\ga^7x^3+\cdots&\ga^7x^4+\cdots&\ga^3x^6+\cdots&\ga^7x^8+\cdots \\
f_2&x+\cdots&\ga^3x^2+\cdots&\ga^7x^3+\cdots&\ga^3x^5+\cdots&\ga^7x^7+\cdots&\ga^5x^8+\cdots
\end{array}
\]
Again \textit{Pairing} and \textit{Voting} steps result in
\[
\begin{array}{ccrr}
\hline
 f_i&g_{i'}&c_i&w_i \\
 \hline
f_0&g_2&3&0\\
f_1&g_0&0&\ga^3\\
f_2&g_1&0&\ga^3
\end{array}
\]
Note that the value $0$ get $3$ votes while the value $\ga^3$ get $0$ votes. Thus $w_{13}=w=0$ is chosen. The pair $f_0,g_2$ is modified by \eqref{fkfmvd}, and the pairs $f_1,g_0$ and $g_2,g_1$ are modified by \eqref{akfjcd}. Thus the Gr\"obner basis with respect to $>_{13}$ for $v^{(13)}=v^{(14)}-\ev(0\cdot x^2y^2)$ is

\[
\begin{array}{r*{6}{|r}}
 &\makebox[50pt][r]{$y^2z$}&\makebox[50pt][r]{$yz$}&\makebox[50pt][r]{$z$}
 &\makebox[50pt][r]{$y^2$}&\makebox[50pt][r]{$y$}&\makebox[50pt][r]{$1$}\\
\hline
g_0&1&\ga^3x+\cdots&\ga^7x^2+\cdots&\ga^3x^4+\cdots&\ga^7x^6+\cdots&\ga^3x^8+\cdots \\
g_1& &x+\cdots&\ga^7x^2+\cdots&2x^5+\cdots&x^7+\cdots&\ga^7x^8+\cdots \\
g_2& & &x+\cdots&\ga^2x^7+\cdots&\ga^5x^8+\cdots&x^9+\cdots \\
f_0& &1&x^2+\cdots& & &\\
f_1&1&x^2+\cdots&\ga^7x^3+\cdots& & &\\
f_2&x+\cdots&\ga^3x^2+\cdots&\ga^7x^3+\cdots&x^4+\cdots&2x^6+\cdots&x^8+\cdots
\end{array}
\]
From the result 
\[
\begin{array}{ccrr}
\hline
f_i&g_{i'}&c_i&w_i \\
\hline
f_0&g_1&2&0\\
f_1&g_2&2&0\\
f_2&g_0&0&2
\end{array}
\]
we set $w_{12}=0$, and the Gr\"obner basis with respect to $>_{12}$ for $v^{(12)}=v^{(13)}-\ev(0\cdot x^3y)$ is
\begin{equation}\label{mskwdd}
\begin{array}{r*{6}{|r}}
 &\makebox[50pt][r]{$y^2z$}&\makebox[50pt][r]{$yz$}&\makebox[50pt][r]{$z$}
 &\makebox[50pt][r]{$y^2$}&\makebox[50pt][r]{$y$}&\makebox[50pt][r]{$1$}\\
\hline
g_0&1&\ga^3x+\cdots&\ga^7x^2+\cdots&\ga^3x^4+\cdots&\ga^7x^6+\cdots&\ga^3x^8+\cdots \\
g_1& &x+\cdots&\ga^7x^2+\cdots&2x^5+\cdots&x^7+\cdots&\ga^7x^8+\cdots \\
g_2& & &x+\cdots&\ga^2x^7+\cdots&\ga^5x^8+\cdots&x^9+\cdots \\
f_0& &1&x^2+\cdots& & &\\
f_1&1&x^2+\cdots&\ga^7x^3+\cdots& & &\\
f_2&x+\cdots&\ga^3x^2+\cdots&\ga^7x^3+\cdots& & &
\end{array}
\end{equation}
For every iteration from this point on, $w_{s}=0$ unanimously, and for the three gaps $5,2,1$, there occurs no modifications. For brevity, we list only voting results:
\[
\begin{array}{ccrr}
&&x^4\\
\hline
 f_i&g_{i'}&c_i&w_i \\
 \hline
f_0&g_0&2&0\\
f_1&g_1&1&0\\
f_2&g_2&2&0
\end{array}
\quad
\begin{array}{ccrr}
&&xy^2\\
\hline
f_i&g_{i'}&c_i&w_i \\
\hline
f_0&g_2&4&0\\
f_1&g_0&1&0\\
f_2&g_1&1&0
\end{array}
\quad
\begin{array}{ccrr}
&&x^2y\\
\hline
f_i&g_{i'}&c_i&w_i \\
\hline
f_0&g_1&3&0\\
f_1&g_2&3&0\\
f_2&g_0&1&0
\end{array}
\quad
\begin{array}{ccrr}
&&x^3\\
\hline
f_i&g_{i'}&c_i&w_i \\
\hline
f_0&g_0&3&0\\
f_1&g_1&2&0\\
f_2&g_2&3&0
\end{array}
\quad
\begin{array}{ccrr}
&&y^2\\
\hline
f_i&g_{i'}&c_i&w_i \\
\hline
f_0&g_2&5&0\\
f_1&g_0&2&0\\
f_2&g_1&2&0
\end{array}
\]
\[
\begin{array}{ccrr}
&&xy\\
\hline
f_i&g_{i'}&c_i&w_i \\
\hline
f_0&g_1&4&0\\
f_1&g_2&4&0\\
f_2&g_0&2&0
\end{array}
\quad
\begin{array}{ccrr}
&&x^2\\
\hline
f_i&g_{i'}&c_i&w_i \\
\hline
f_0&g_0&4&0\\
f_1&g_1&3&0\\
f_2&g_2&4&0
\end{array}
\quad
\begin{array}{ccrr}
&&y\\
\hline
 f_i&g_{i'}&c_i&w_i \\
 \hline
f_0&g_1&5&0\\
f_1&g_2&5&0\\
f_2&g_0&3&0
\end{array}
\quad
\begin{array}{ccrr}
&&x\\
\hline
f_i&g_{i'}&c_i&w_i \\
\hline
f_0&g_0&5&0\\
f_1&g_1&4&0\\
f_2&g_2&5&0
\end{array}
\quad
\begin{array}{ccrr}
&&1\\
\hline
f_i&g_{i'}&c_i&w_i \\
\hline
f_0&g_0&6&0\\
f_1&g_1&5&0\\
f_2&g_2&6&0
\end{array}
\]
Thus \eqref{mskwdd}, with no modifications, remains as a Gr\"obner basis with respect to $>_{-1}$. Hence the recovered message is 
\[
	(w_{0},w_{3},w_{4},w_{6},w_{7},w_{8},w_{9},w_{10},w_{11},w_{12},w_{13},w_{14},w_{15},w_{16})=0\in\F^{14}
\]
and the recovered codeword is the zero codeword.
\section{Final Remarks}\label{sec_dwkqqw}

We presented a unique decoding algorithm based on interpolation. Like the syndrome decoding algorithm, our decoding algorithm corrects errors of up to half of the order bound. It computes the message directly from the received vector under evaluation encoding, which is a distinctive feature of list decoding. Like K\"otter's algorithm for syndrome decoding, our decoding algorithm is amenable to a parallel hardware architecture. 

We would like to thank the anonymous referees for thoughtful comments that much improved the original paper.



\end{document}